\documentclass[twocolumn,preprintnumbers,amsmath,amssymb,prb]{revtex4}
\usepackage{graphicx}
\usepackage{bm}
\usepackage{color}

\begin{document}
\title{BCS-BEC Crossover in the Two-Dimensional Attractive Hubbard Model:\\ Variational Cluster Approach}

\author{Tatsuya Kaneko and Yukinori Ohta}
\affiliation{Department of Physics, Chiba University, Chiba 263-8522, Japan}

\received{September 25, 2013}
\begin{abstract}
We use the variational cluster approximation to study the superconducting ground state in the two-dimensional attractive Hubbard model, putting particular emphasis on the significance of quantum fluctuations of the system. 
We first show that the order parameter is suppressed in comparison with that obtained using the mean-field theory owing to the effects of spatial fluctuations in two-dimensional systems.
We then show that the calculated Bogoliubov quasiparticle spectra and condensation amplitude clearly exhibit the character of Cooper pairs in momentum space and that the pair coherence length $\xi$ evaluated from the condensation amplitude demonstrates a smooth crossover in real space from a weakly paired BCS state ($\xi\gg a$) to a BEC state of tightly bound pairs ($\xi\ll a$), where $a$ is the lattice constant.  
The calculated kinetic and potential energies in the superconducting and normal ground states indicate that the superconducting state in the weak-coupling region is driven by the gain in potential energy, while that in the strong-coupling region is driven by the gain in kinetic energy. 
\end{abstract}

\maketitle

\section{Introduction}

The physics of the crossover between the Bardeen-Cooper-Schrieffer (BCS) state  of weakly bound Cooper pairs and the Bose-Einstein condensed (BEC) state of tightly bound composite bosons, i.e., the BCS-BEC crossover, has long been one of the major issues in condensed matter physics.  
The idea of a continuous crossover between the BCS and BEC limits first arose in the 1960s as a problem of exciton condensation near the semimetal-semiconductor transition, which is called the `excitonic insulator' state.\cite{keldysh,kozolov,halperin1}  
The BCS-BEC crossover also attracted attention from the early stages of the theory of superconductivity, where Eagles\cite{eagles} first addressed this issue in metals with a very low electron density.  
In 1980, using a variational approach, Leggett\cite{leggett} showed a smooth crossover from the weak-coupling BCS state to the strong-coupling BEC state at zero temperature.
The critical temperature $T_c$ across the BCS-BEC crossover was first evaluated by Nozi\`eres and Schmitt-Rink.\cite{nozieres}  
In 1986, the discovery of high-$T_c$ cuprate materials, where the coherence length is only a few times larger than the lattice spacing, led to intensive discussion on the possible realization of the BCS-BEC crossover in cuprate superconductors.\cite{leggett2}  


In systems of ultracold fermionic atoms, the crossover between the BCS-type and BEC-type superfluid states has also been observed,\cite{jochim,zwierlein1,greiner,bourdel,regal,bartenstein,zwierlein2,bloch} where the interaction strength is controlled through a magnetically tuned Feshbach resonance. 
In particular, a two-dimensional ultracold Fermi gas has recently been realized experimentally in a very controlled manner,\cite{martiyanov,dyke,frohlich,feld,sommer,zhang} although the inclusion of an optical lattice potential to realize the two-dimensional Fermi-Hubbard model has yet to be carried out.\cite{feld}   
In the study of such two-dimensional Fermi gas systems, where tunable and clean samples have become available,\cite{fischer} the Berezinskii-Kosterlitz-Thouless superfluidity\cite{hadzibabic} without condensation,\cite{frohlich,wzhang} the presence of the pairing pseudogap,\cite{feld} BCS-BEC crossover,\cite{bertaina,fischer} and two- to three-dimensional crossovers\cite{dyke,sommer} have been discussed intensively in recent years.

Motivated by the above developments in the field, in this study, we investigate the BCS-BEC crossover of the superconducting ground state in the two-dimensional attractive Hubbard model.\cite{micnas}  
Thus far, focusing on numerical studies, the BCS-BEC crossover of this model has been explored mostly using the dynamical mean-field theory (DMFT),\cite{capone,garg,toschi,bauer,koga} where the correlation effects can be taken into account only in the 
infinite dimension.  
The cellular DMFT somehow improves the effects of finite dimensionality.\cite{kyung}  
We here employ the variational cluster approximation (VCA) based on the self-energy functional theory (SFT),\cite{potthoff1} where we can take into account the effects of short-range spatial correlations even in low-dimensional systems, thereby reproducing the momentum dependences of physical quantities precisely.  
This method has been shown to be useful for discussing the spontaneous symmetry breaking of correlated electron models beyond the mean-field theory;\cite{dahnken,seki,kaneko} however, to the best of our knowledge, it has not been used to study the BCS-BEC crossover in the attractive Hubbard model.  

Therefore, using VCA, we first discuss the interaction ($U$) dependence of the order parameter of superconductivity.  
We will show that the order parameter is suppressed in comparison with that obtained using the mean-field theory owing to the effects of spatial fluctuations in low-dimensional systems.
In order to show the dynamics of the BCS-BEC crossover, we will then use the cluster perturbation theory (CPT)\cite{senechal1} to calculate the single-particle and anomalous Green's functions.  
We will present the single-particle spectra and densities of states to clarify the behavior of the superconducting gap.  
We will also present the Bogoliubov quasiparticle spectra and condensation amplitude to discuss the character of Cooper pairs in the BCS and BEC states.  
In particular, we will  evaluate the pair coherence length $\xi$ from the condensation amplitude and demonstrate the smooth crossover from a weakly paired BCS state ($\xi \gg a$) to a BEC state of tightly bound pairs ($\xi \ll a$), where $a$ is the lattice constant.  
We will finally calculate the kinetic and potential energies in the superconducting and normal ground states and show that the superconducting state is driven by the gain in potential energy in the BCS state, but by the gain in kinetic energy in the BEC state, in agreement with previous theories.\cite{toschi,kyung}

This paper is organized as follows.  In Sect.~2, we will introduce the attractive Hubbard model and briefly summarize the method of VCA for discussing an $s$-wave superconducting state.  
In Sect.~3, we will show the calculated results for various physical quantities and discuss the BCS-BEC crossover.  
A  summary of the paper will be given in Sect.~4. 
  
\section{Model and Method}

To discuss the BCS-BEC crossover, we use the attractive Hubbard model defined as
\begin{equation}
\mathcal{H}=-t\sum_{\langle i,j\rangle,\sigma}c^{\dag}_{i\sigma}c_{j\sigma}
-U \sum_{i}n_{i\uparrow}n_{i\downarrow}-\mu\sum_{i,\sigma}n_{i\sigma}, 
\end{equation}
where $c^{\dag}_{i\sigma}$ ($c_{i\sigma}$) is the fermion creation (annihilation) operator with spin $\sigma(=\uparrow,\downarrow)$ at site $i$ and $n_{i\sigma}=c^\dag_{i\sigma}c_{i\sigma}$.  
$t$ is the hopping integral between nearest-neighbor sites, $U\:(>0)$ is the on-site attractive interaction, and $\mu$ is the chemical potential for maintaining the number of particles in the system.  
It is known\cite{scalettar,moreo} that the superconducting state is always realized in two and higher dimensions at $T=0$ K for all values of $U$ $(>0)$ and in the entire particle density range, except at half filling where the superconducting and density-wave states are degenerate.  

We use VCA, which is an extension of the CPT based on the self-energy functional theory (SFT).  
Following Potthoff,\cite{potthoff2} we write the grand-potential functional as $\Omega[\bm{\Sigma}]=\Lambda[\bm{\Sigma}]-{\rm Tr}\ln(-{\bm G}^{-1}_{0}+\bm{\Sigma})$, where $\Lambda[\bm{\Sigma}]$ is the Legendre transform of the Luttinger-Ward functional and $\bm{G}_0$ is the noninteracting Green's function.  
We call $\bm{\Sigma}$ the trial self-energy and the stationary condition 
\begin{equation}
\frac{\delta\Omega[\bm{\Sigma}]}{\delta\bm{\Sigma}}=0 
\end{equation}
gives the Dyson equation; at this stationary point, this functional gives the grand potential of the system.  
The SFT provides a way to compute $\Omega$ based on the fact that the functional form of $\Lambda[\bm{\Sigma}]$  
depends only on the interaction terms of the Hamiltonian.  
Here, we introduce disconnected finite-size clusters forming a superlattice as a reference system; each cluster has the exact grand potential $\Omega'=\Lambda[\bm{\Sigma}']-{\rm Tr}\ln(-\bm{G}'^{-1}_{0}+\bm{\Sigma}')$, 
where $\bm{\Sigma}'$ is the exact self-energy of the reference system.  
Because the original and reference systems have the same interaction term, their functional forms $\Lambda[\bm{\Sigma}]$ are the same.  
Therefore, by restricting the trial $\bm{\Sigma}$ to $\bm{\Sigma}'$, we can eliminate the functional $\Lambda[\bm{\Sigma}]$ and obtain 
\begin{equation}
\Omega[\bm{\Sigma}']=\Omega'-{\rm Tr}\ln(\bm{I}-\bm{VG'}), 
\end{equation}
where $\bm{I}$ is the unit matrix, $\bm{V}=\bm{G}'^{-1}_{0}-\bm{G}^{-1}_{0}$, and $\bm{G}'=(\bm{G}'^{-1}_{0}-\bm{\Sigma}')^{-1}$ is the exact Green's function of the reference system.  

The trial self-energy for the variational method is generated from the exact self-energy (or the exact Green's function) of the reference system, for which the Hamiltonian is defined as 
\begin{align}
&\mathcal{H}'=\mathcal{H}+\mathcal{H}_\mathrm{pair}+\mathcal{H}_\mathrm{local} \\
&\mathcal{H}_\mathrm{pair}=\Delta'\sum_{i}(c^{\dag}_{i\uparrow}c^{\dag}_{i\downarrow}+\mathrm{H.c.}) \\
&\mathcal{H}_\mathrm{local}= \varepsilon'\sum_{i,\sigma} n_{i\sigma}, 
\end{align} 
where the Weiss field for the $s$-wave pairing $\Delta'$ and the on-site potential $\varepsilon'$ are variational parameters.  
Note that $\varepsilon'$ is introduced in order to calculate the particle density $n$ correctly.  
Then, we solve the ground-state eigenvalue problem $\mathcal{H}'|\psi_0\rangle = E_0|\psi_0\rangle$ of a finite-size ($L_c$ sites) cluster and calculate the trial Green's function 
by the Lanczos exact-diagonalization method.  
We use the Nambu formalism $\varPsi^{\dag}_{i} = (c^{\dag}_{i\uparrow},c_{i\downarrow})$ to solve the eigenvalue problem of Eq.~(4); the Green's function matrix in Eq.~(3) is then defined as
\begin{align}
\hat{\bm{G}'}(\omega )
=\left( \begin{array}{cc}
\bm{G}'(\omega ) &\bm{F}'(\omega ) \\
\bm{F}'^{\dag}(\omega )  &-\bm{G}'(-\omega )  \\
\end{array} \right), 
\end{align}
where $\bm{G}'$ and $\bm{F}'$ are the $L_c \times L_c$ matrices, and each matrix element is defined as $G'_{ij}(\omega)=\langle\langle c_{i\uparrow}; c^{\dag}_{j\uparrow}\rangle\rangle_{\omega}$ and $F'_{ij}(\omega)=\langle\langle c_{i\uparrow}; c_{j\downarrow}\rangle\rangle_{\omega}$, respectively.  
We will denote all the Nambu matrices by a `hat' on top.  
The matrix $\bm{V}$ in Eq.~(3) is given as 
\begin{align}
\hat{\bm{V}}(\bm{K})=\left(
\begin{array}{cc}
\bm{T}(\bm{K})-\varepsilon'\bm{I}&-\Delta'\bm{I}\\
-\Delta'\bm{I}&-\bm{T}(\bm{K})+\varepsilon'\bm{I}\\
\end{array} 
\right),
\end{align}
where $\bm{T}(\bm{K} )$ is the intercluster hopping matrix with $T_{ij}(\bm{K} )=-t \sum_{\bm{X},x}e^{i\bm{K}\cdot\bm{X}}\delta _{i+x,j}\delta_{\bm{R}+\bm{X},\bm{R}'}$, where $x$ denotes the neighboring site of the $i$-th site and $\bm{X}$ denotes the neighboring cluster of the $\bm{R}$-th cluster. 

Using the matrices $\hat{\bm{G}}$ and $\hat{\bm{V}}$, we can evaluate the functional
\begin{align}
\Omega=\Omega'-\frac{1}{N}\oint_{C}\frac{{\rm d}z}{2\pi i}\sum_{\bm{K}}\ln\det
\left[\hat{\bm{I}}-\hat{\bm{V}}(\bm{K})\hat{\bm{G}}'(z)\right], 
\end{align}
where the $\bm{K}$-summation is performed in the reduced Brillouin zone of the superlattice and the contour $C$ of the frequency integral encloses the negative real axis.
The variational parameters $\Delta'$ and $\varepsilon'$ are optimized on the basis of the variational principle, i.e., $(\partial\Omega/\partial\Delta',\partial\Omega/\partial\varepsilon')=(0,0)$.  
The solution with $\Delta'\ne 0$ corresponds to the superconducting state.  
The average particle density $n$ $(=\langle n_{i\sigma}\rangle)$ is expressed as
\begin{align}
n=\frac{1}{NL_c}\oint_{C}\frac{{\rm d}z}{2\pi i}\sum_{\bm{K}}\sum^{L_c}_{i=1}\mathcal{G}_{ii}(\bm{K},z) 
\end{align}
and, throughout the paper, the chemical potential $\mu$ is determined to maintain the particle density $n$ at quarter filling, $2n=\langle n_{i\uparrow}\rangle+\langle n_{i\downarrow}\rangle=0.5$.  
$\bm{\mathcal{G}}$ in Eq.~(10) is the diagonal term ($L_c \times L_c$ matrix) of $\hat{\bm{\mathcal{G}}}(\bm{K},\omega)=\big[\hat{\bm{G}}'^{-1}(\omega)-\hat{\bm{V}}(\bm{K})\big]^{-1}$.  

A cluster of size $L_{c}=2\times 2=4$ is used as a reference system; the effects of on-site correlations within this cluster are taken into account exactly.  
Detailed techniques of VCA can be found in Refs.\cite{potthoff2,senechal2}.  

\begin{figure}[htb]
\begin{center}
\resizebox{8.1cm}{!}{\includegraphics{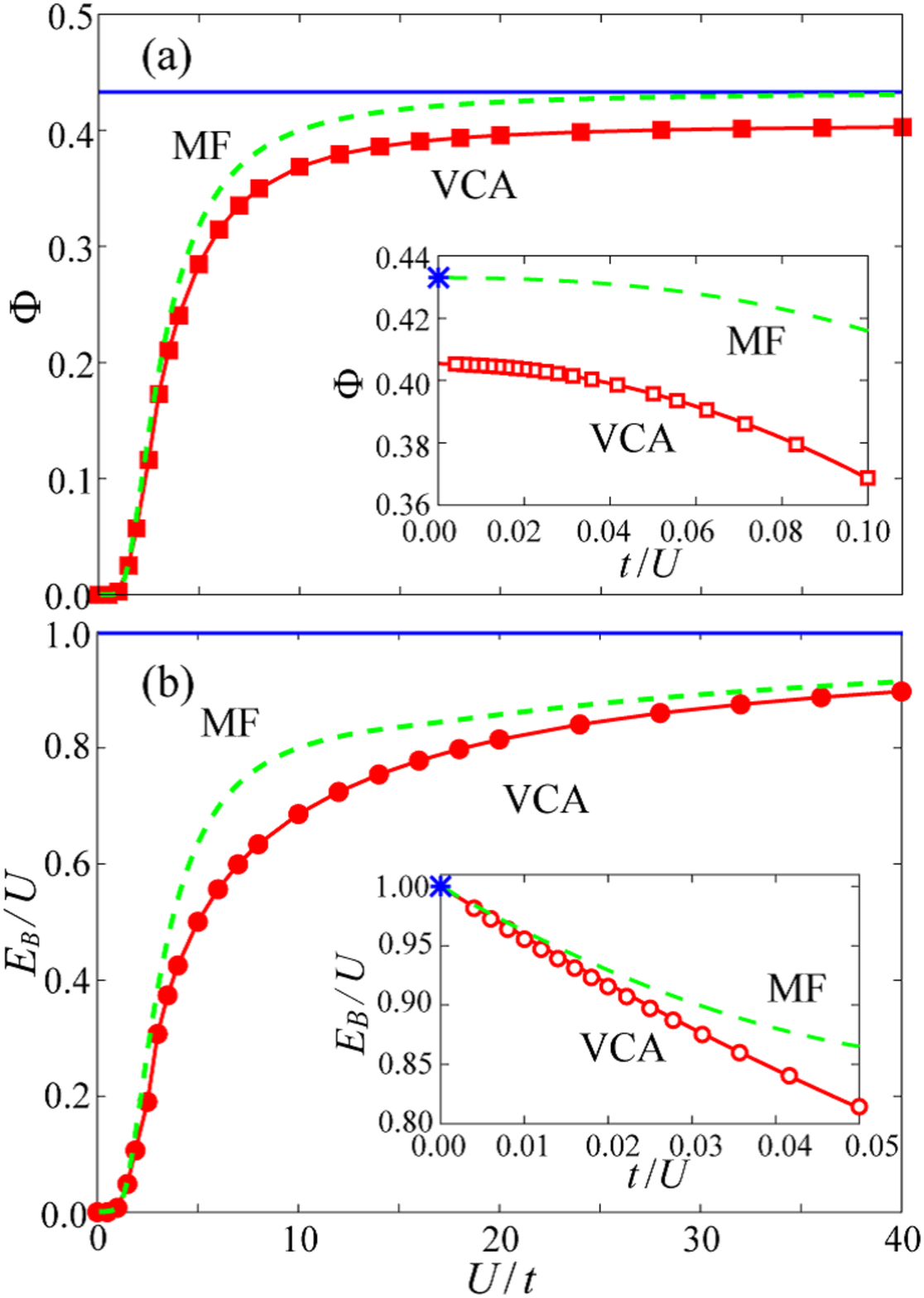}}
\caption{(Color online) (a) Anomalous expectation values 
$\Phi=\langle c_{i\downarrow}c_{i\uparrow}\rangle$ 
calculated using VCA (squares) and the MF theory (dashed line) 
as a function of $U/t$ at quarter filling ($n=0.25$).  
The horizontal line indicates the $\Phi $ in the MF theory 
in the strong-coupling limit, $\Phi_{\mathrm{MF}}=\sqrt{3}/4$.  
The inset shows the $\Phi$ values calculated using VCA (open squares) and MF theory (dashed line) 
in the strong-coupling region as a function of $t/U$.
(b) Binding energies of a pair $E_B/U$ calculated using VCA (circles) and the MF theory (dashed line) 
as a function of $U/t$ at quarter filling ($n=0.25$).  
The horizontal line indicates the $E_B/U$ in the MF theory 
in the strong-coupling limit, $E_B=U$.  
The inset shows the $E_B/U$ values calculated using VCA (open circles) and the MF theory (dashed line) 
in the strong-coupling region as a function of $t/U$.
}  
\label{fig1}
\end{center}
\end{figure}

\begin{figure*}[!t]
\begin{center}
\resizebox{15cm}{!}{\includegraphics{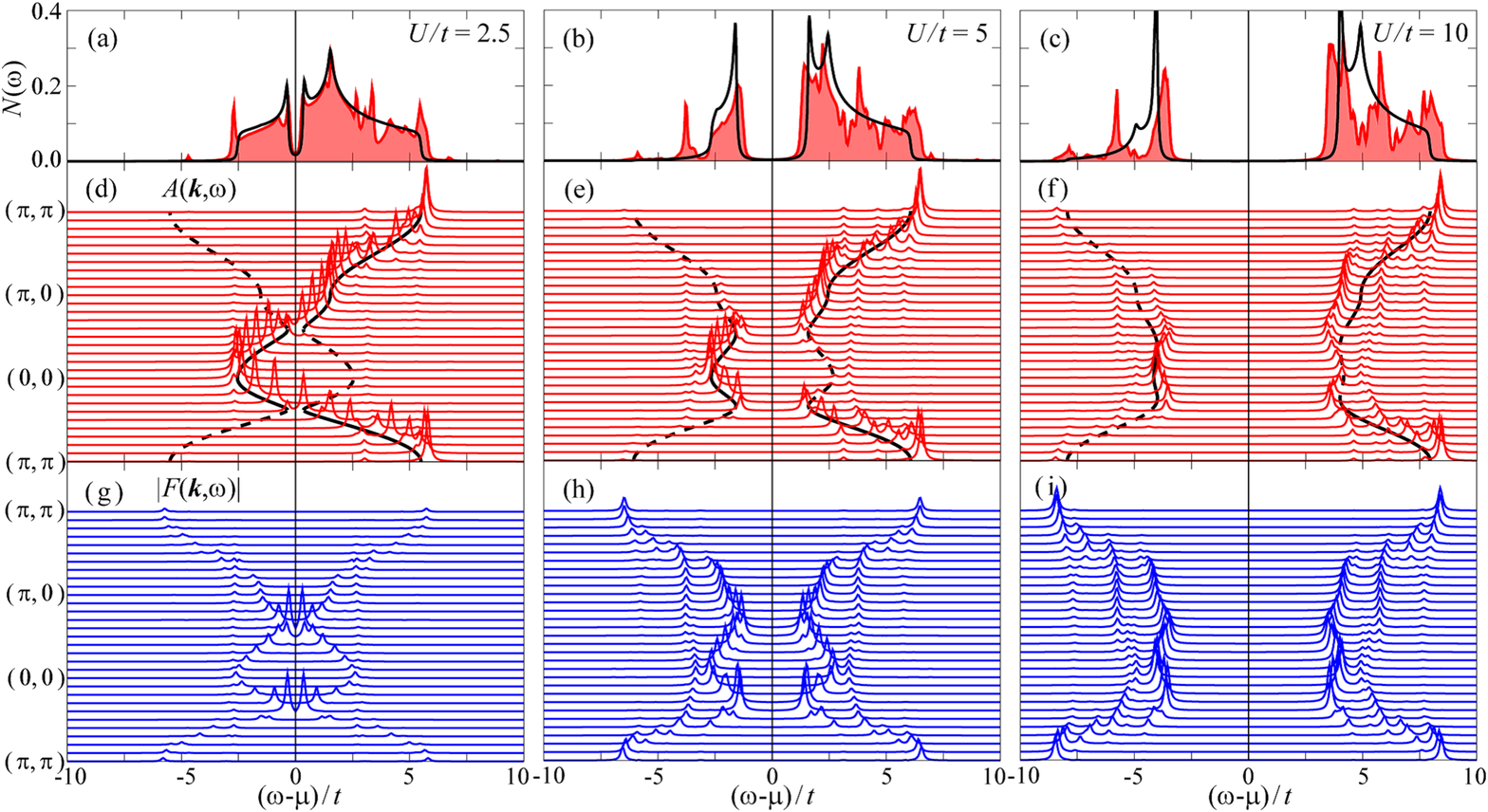}}
\caption{(Color online) 
(a)-(c) Densities of states $N(\omega)$, 
(d)-(f) single-particle spectra $A(\bm{k},\omega)$, and 
(g)-(i) Bogoliubov quasiparticle spectra $F(\bm{k},\omega)$ 
calculated at $U/t=2.5$ (left), $U/t=5$ (center), and $U/t=10$ (right) 
at quarter filling ($n=0.25$).  
The densities of state and quasiparticle dispersions evaluated in the 
MF theory (solid and dashed lines) are also shown in (a)-(c) and (d)-(f), 
respectively.
The Lorentzian broadening of $\eta/t=0.05$ is used for $N(\omega)$, and 
$\eta/t=0.1$ is used for $A(\bm{k},\omega)$ and $F(\bm{k},\omega)$.} 
\label{fig2}
\end{center}
\end{figure*}

\section{Results of Calculation}
\subsection{Order parameter}

We first calculate the $U$ dependence of the superconducting order parameter $\Delta=U\langle c_{i\downarrow}c_{i\uparrow}\rangle$.  
Within the framework of VCA, the anomalous expectation value $\Phi=\langle c_{i\downarrow}c_{i\uparrow}\rangle $ is defined as 
\begin{align}
\Phi=\frac{1}{NL_c}\oint_{C}\frac{{\rm d}z}{2\pi i}\sum_{\bm{K}}\sum^{L_c}_{i=1}\mathcal{F}_{ii}(\bm{K},z), \label{eq11}
\end{align}
where $\bm{\mathcal{F}}$ is the off-diagonal term of the Green's function $\hat{\bm{\mathcal{G}}}(\bm{K},\omega)$.  
We also evaluate the binding energy of the pair $E_B$ from the single-particle excitation gap.
For comparison with the results of VCA, we also evaluate the order parameter in the mean-field (MF) theory, which gives $\Delta$ as a solution of the self-consistent equations.  

The results for $\Phi$ and $E_B$ calculated using the VCA and MF theory are shown in Fig.~\ref{fig1}.  
In the MF theory, the order parameter $\Delta^{\mathrm{MF}}$ increases exponentially with $U$, thereby satisfying the relation $E^{\mathrm{MF}}_B = 2\Delta^{\mathrm{MF}}$ in the weak-coupling limit.  
In the strong-coupling limit, on the other hand, $\Delta^{\mathrm{MF}}=U\sqrt{n(1-n)}=\sqrt{3}U/4$ $(\Phi^{\mathrm{MF}}=\sqrt{3}/4)$ and $E^{\mathrm{MF}}_B = U$ at $n=0.25$, regardless of the spatial dimension. 
We find that the result of VCA exhibits the same behavior as that of the MF theory in the weak-coupling limit: $\Delta$ increases exponentially with $U$, satisfying the relation $E_B = 2\Delta$, which recovers the exponential behavior of the BCS mean-field theory.  
In the intermediate-coupling region, we find that $\Phi$ and $E_B$ are significantly suppressed in comparison with 
those in the case of the MF theory, which is due to the quantum fluctuations of the system. 
In the strong-coupling limit, $E_B$ converges to the result of the MF theory, $E_B=E^{\mathrm{MF}}_B=U$ [see the inset of Fig.~1(b)], but $\Phi$ is suppressed in comparison with the result of the MF theory, $\Phi \sim 0.405 < \Phi ^{\mathrm{MF}}$ at $U \rightarrow \infty$ [see the inset of Fig.~1(a)].

In the strong-coupling limit, the attractive Hubbard model can be mapped onto the spin-1/2 Heisenberg model in a magnetic field, 
\begin{align}
\mathcal{H}_{eff} = J\sum_{\langle i,j\rangle} \bm{S}_i\cdot\bm{S}_j-h\sum_{i}S^z_i, 
\end{align}
where we use the particle-hole transformation $a_{i\uparrow}=c_{i\uparrow}$ and $a_{i\downarrow}=(-1)^ic^{\dag}_{i\downarrow}$,\cite{shiba} and define $\bm{S}_i=\frac{1}{2}\sum a^{\dag}_{i\alpha}\sigma_{\alpha\beta}a_{i\beta}$, $J=4t^2/|U|$, and $h=2\mu+|U|$.  
The superconducting state in the original model at quarter filling ($n=0.25$) corresponds to the antiferromagnetically ordered state in the $xy$ plane in the effective model with the magnetization $m = \sum\langle S^z_i \rangle /N = 0.25$.  
It is known\cite{scalettar,luscher} that, in the two-dimensional square lattice, strong quantum fluctuations caused by the low dimensionality of the system suppress the long-range staggered magnetic order in the $xy$ plane in comparison with those in the case of classical approximation.  
Therefore, because VCA takes into account the short-range spatial correlations and quantum fluctuations in the low-dimensional systems, it is reasonable that the order parameter $\Phi$ obtained using VCA is significantly suppressed in comparison with the result of the MF theory.

To compare the result of VCA with those of the DMFT calculations, which are justified in the infinite dimension, we notice that our result for $\Phi$ in the strong-coupling limit is quite different: $\Phi$ in DMFT increases to the constant value obtained in the MF theory,\cite{garg,bauer} whereas, in VCA, it converges to a significantly smaller value, as shown in Fig.~\ref{fig1}(a).  
The MF theory for the Heisenberg model is exact in the infinite dimension; therefore, the results of the DMFT calculations are consistent with the results of the MF theory in the strong-coupling limit.  
The difference between the DMFT and VCA results is thus caused by the effects of spatial quantum fluctuations in  low-dimensional systems, which the DMFT cannot take into account.  

\begin{figure*}[htb]
\begin{center}
\resizebox{15cm}{!}{\includegraphics{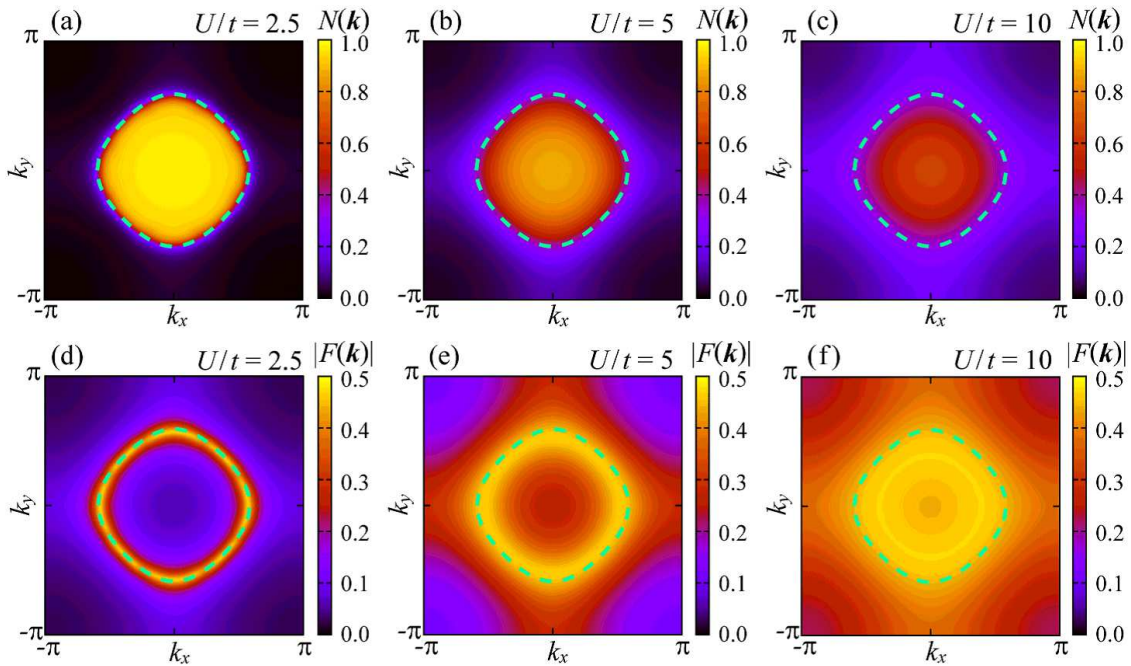}}
\caption{(Color online) 
(a)-(c) Fermion momentum distribution function $N(\bm{k})$ and 
(d)-(f) condensation amplitude $F(\bm{k})$ calculated 
at $U/t=2.5$ (left), $U/t=5$ (center), and $U/t=10$ (right)
at quarter filling ($n=0.25$).  
The dashed line indicates the Fermi momentum.
}
\label{fig3}
\end{center}
\end{figure*}

\subsection{Spectra and momentum distributions}

The single-particle and anomalous Green's functions are calculated using CPT with the optimized variational parameters, which are defined as
\begin{align}
\mathcal{G}_{\mathrm{cpt}}(\bm{k},\omega)=\frac{1}{L_c}\sum^{L_c}_{i,j=1}\mathcal{G}_{ij}(\bm{k},\omega)e^{-i\bm{k}\cdot(\bm{r}_i-\bm{r}_j)} \\
\mathcal{F}_{\mathrm{cpt}}(\bm{k},\omega)=\frac{1}{L_c}\sum^{L_c}_{i,j=1}\mathcal{F}_{ij}(\bm{k},\omega)e^{-i\bm{k}\cdot(\bm{r}_i-\bm{r}_j)},
\end{align}
from which we calculate the single-particle and Bogoliubov quasiparticle spectra defined respectively as 
\begin{align}
&A(\bm{k},\omega)=-\frac{1}{\pi}\mathrm{Im}\; \mathcal{G}_{\mathrm{cpt}}(\bm{k},\omega+i\eta) \\
&F(\bm{k},\omega)=-\frac{1}{\pi}\mathrm{Im}\;  \mathcal{F}_{\mathrm{cpt}} (\bm{k},\omega+i\eta), 
\end{align}
where $\eta$ is the artificial Lorentzian broadening.  
We also calculate the density of states defined as 
\begin{align}
N(\omega) = \frac{1}{N} \sum_{\bm{k}} A(\bm{k},\omega).
\end{align}

In Fig.~\ref{fig2}, we show the calculated results for $A(\bm{k},\omega)$, $F(\bm{k},\omega)$, and $N(\omega)$ from the weak-coupling region to the strong-coupling region.  
In the weak-coupling region (at $U/t=2.5$), $A(\bm{k},\omega)$ [or $N(\omega)$] shows a tiny superconducting gap at the Fermi momentum $\bm{k}_\mathrm{F}$, together with coherence peaks at the edges of the gap, indicating 
the existence of weakly bound Cooper pairs.
The gap width and peaks of $A(\bm{k},\omega)$ are consistent with quasiparticle spectra in the MF theory.  
Note that the Fermi momentum $\bm{k}_\mathrm{F}$ is defined as $\varepsilon_{\bm{k}_\mathrm{F}}=\mu$ (at $U=0$), where $\varepsilon_{\bm{k}}=-2t(\cos k_x+\cos k_y)$.  
$F(\bm{k},\omega)$ has a sharp peak at $\bm{k}_\mathrm{F}$ and its intensity rapidly decreases as the momentum goes 
away from $\bm{k}_\mathrm{F}$.  
With increasing $U$, each of the pairs becomes more strongly bound and the superconducting gap becomes larger.  
In accordance with the $E_B$ shown Fig.~1(b), $N(\omega)$ exhibits a spectral gap that is suppressed in comparison with the results of the MF theory, as shown in Figs.~2(b) and 2(c). 
At $U/t=5$, $F(\bm{k},\omega)$ still has a strong peak at approximately $\bm{k}_\mathrm{F}$. In comparison with the spectra at $U/t=2.5$, $F(\bm{k},\omega)$ has strong peaks even as the momentum goes away from $\bm{k}_\mathrm{F}$.  
In the strong-coupling region (at $U/t=10$), the spectra show a large superconducting gap, and the peaks of $F(\bm{k},\omega)$ spread out over the entire Brillouin zone.

In order to see the character of Cooper pairs in more detail in the momentum space, we also calculate the fermion momentum distribution function and condensation amplitude, which are defined respectively as
\begin{align}
N({\bm k})&= \oint_{C}
\frac{{\rm d}z}{2\pi i}
\mathcal{G}_{\mathrm{cpt}}({\bm k},z), \\
F({\bm k})&=\oint_{C}
\frac{{\rm d}z}{2\pi i}
\mathcal{F}_{\mathrm{cpt}}({\bm k},z).  
\end{align}
The calculated results for $N(\bm{k})$ and $F(\bm{k})$ are shown in Fig.~\ref{fig3}.  In the weak-coupling region (at $U/t=2.5$), $N(\bm{k})$ shows the typical form known from the BCS theory, i.e., momenta inside of $\bm{k}_\mathrm{F}$ are mostly occupied ($N(\bm{k})\simeq 1$) and $N(\bm{k})$ slightly broadens at $\bm{k}_\mathrm{F}$, dropping from 1 to 0 over the energy scale of the order parameter.  
Corresponding to $N(\bm{k})$, $F({\bm{k}})$ exhibits a sharp peak at $\bm{k}_\mathrm{F}$ ($|F(\bm{k}_\mathrm{F})|\simeq 0.5$) and decreases rapidly as the momentum goes away from $\bm{k}_\mathrm{F}$.  
The sharp peak of $F(\bm{k})$ in the $\bm{k}$-space indicates that the radius of the pair is large in real space (weakly bound pairs).  
With increasing $U$, $N({\bm{k}})$ and $F({\bm{k}})$ become broader in momentum space, indicating that the radius of the pair becomes smaller in real space.
In the strong-coupling region (at $U/t=10$),  $F({\bm{k}})$ is spread out over the Brillouin zone; therefore, the pairs are tightly bound in real space.

\subsection{Pair coherence length}

In order to see the spatial extension of the Cooper pair directly, 
we evaluate the pair coherence length $\xi$ defined as
\begin{align}
\xi^2  =\frac{\sum_{\bm{r}}\bm{r}^2|F(\bm{r})|^{2}}{{\sum_{\bm{r}}|F(\bm{r})|^{2}}}
 = \frac{\sum_{\bm{k}}|\bm{\nabla}_{\bm{k}} F(\bm{k})|^{2}}{\sum_{\bm{k}}|F(\bm{k})|^{2}},
\end{align}
where $F(\bm{r})=\frac{1}{\sqrt{L}}\sum_{\bm{r}'}\langle c_{\bm{r}'+\bm{r}\downarrow}c_{\bm{r}'\uparrow}\rangle$ is the condensation amplitude for a Cooper pair with a distance $\bm{r}$ in real space.\cite{ohta} 
The $\bm{k}$-summation was performed with $500\times500$ $\bm{k}$ points in the first Brillouin zone.

\begin{figure}[!tb]
\begin{center}
\resizebox{8.5cm}{!}{\includegraphics{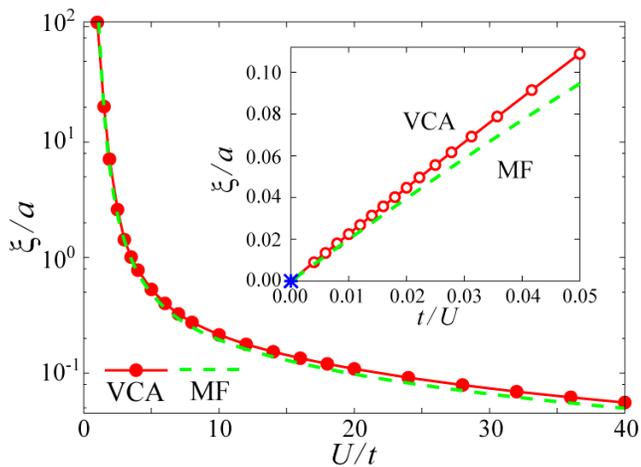}}
\caption{(Color online) Pair  coherence length $\xi/a$ 
 calculated using VCA (circles) and MF theory (dashed line) as a function of $U/t$ for quarter filling ($n=0.25$).
The inset shows $\xi/a$ calculated using VCA (open circles) and MF theory (dashed line) in the strong-coupling limit.}
\label{fig4}
\end{center}
\end{figure}

In Fig.~\ref{fig4}, we show the results for $\xi$ calculated using VCA and the MF theory.  
We find that, corresponding to the calculated results for $F(\bm{k})$ [see Figs.~\ref{fig3}(d)-\ref{fig3}(f)], the pair  coherence length $\xi$ is much larger than the lattice constant $a$ in the weak-coupling region.  
With increasing $U$, $\xi$ decreases smoothly to much smaller values than the lattice constant in the strong-coupling region, indicating that a smooth crossover occurs from the weakly paired BCS-like state ($\xi \gg a$) to the BEC state of tightly bound pairs ($\xi \ll a$).  
Note that $\xi$ is already of the size of the lattice constant at $U/t\sim 3.5$.  
In comparison with the results of VCA and the MF theory, $\xi$ evaluated by VCA is significantly larger than the results of the MF theory, which is due again to the quantum fluctuations of the system, just as in $E_B$ shown 
in Fig.~1(b).  
In the strong-coupling limit ($U\rightarrow \infty$), $\xi$ evaluated using VCA and the MF theory converges to 0 ($\xi \rightarrow 0$).

\subsection{Ground-state energy}
\begin{figure}[!t]
\begin{center}
\resizebox{8.25cm}{!}{\includegraphics{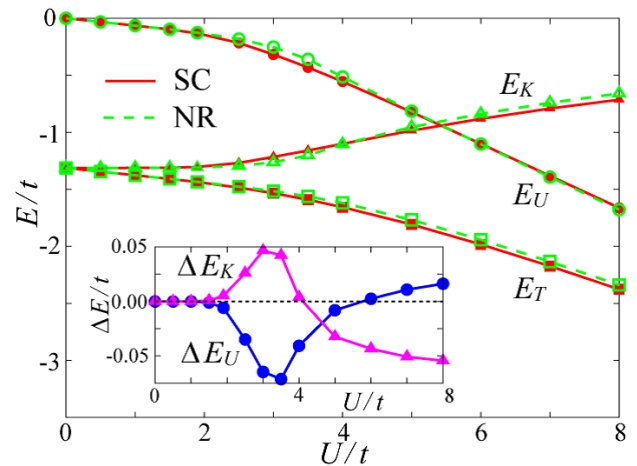}}
\caption{(Color online) 
Calculated total energy $E_{T}$ (squares), kinetic energy 
$E_{K}$ (triangles), and potential energy $E_{U}$ (circles) 
of the superconducting (SC) and normal (NR) ground states as a 
function of $U/t$ at quarter filling ($n=0.25$).  
The inset shows the energy differences between the 
superconducting and normal ground states.}
\label{fig5}
\end{center}
\end{figure}

Finally, we calculate the ground-state energies of the attractive Hubbard model in the superconducting and normal states.  
The total ground-state energy $E_T$ is given as $E_T =\Omega+2\mu n$.  
Using the double occupancy defined as $D_{occ}=\langle  n_{i\uparrow}n_{i\downarrow} \rangle = dE_T/dU$, we obtain the potential energy $E_U$ as $E_U = -UD_{occ}$, and the kinetic energy $E_K$ as $E_K = E_T-E_U$.  
The calculated results are shown in Fig.~\ref{fig5}, where the difference $\Delta E$ denotes the energy of the superconducting state minus the energy of the normal state.  

First, let us  consider the behavior of $E_U$ and $E_K$.  
In the noninteracting limit, $D_{\mathrm{occ.}}$ is given by $n^2=0.0625$ since $\langle n_{i\uparrow}n_{i\downarrow}\rangle = \langle n_{i\uparrow} \rangle\langle n_{i\downarrow} \rangle = n^2$ 
in the uncorrelated fermion systems.  
In the strong-coupling limit, on the other hand, all the fermions are tightly bound to form composite bosons and hence $D_{\mathrm{occ.}}$ is given by the particle density as $D_{\mathrm{occ.}}=n=0.25$.  
The calculated result for $D_{\mathrm{occ.}}$ shows a smooth crossover from the weakly paired BCS state ($D_{\mathrm{occ.}}\simeq n^2$) to the tightly paired BEC state ($D_{\mathrm{occ.}}\simeq n$).  
Therefore, with increasing $U$, $E_U$ decreases owing to the pair formation and $E_K$ increases owing to the gap opening, resulting in a gradual decrease in $E_T$, irrespective of whether the ground state is superconducting or normal; the effects of the presence of the order parameter are found to be rather small.  

Then, let us see the effects of the order parameter, the results of which are shown in the inset of Fig.~\ref{fig5}.  
In the weak-coupling region, we find that the superconducting state occurs owing to the loss of kinetic energy ($\Delta E_{K}>0$) and gain in potential energy ($\Delta E_U < 0$).  
This means that, in the BCS weak-coupling limit, the decrease in the potential energy due to the Cooper pair 
formation overwhelms the loss of kinetic energy due to the broadening of the Fermi edge.  
Thus, in the weak-coupling region, the superconducting state is driven by the gain in potential energy.  
In our calculation, the BCS features ($\Delta E_K>0$ and $\Delta E_U<0$) vanish at $U/t \sim 4$.  

In the strong-coupling region, on the other hand, the roles are interchanged, i.e., the superconducting state is characterized by $\Delta E_{K}<0$ and $\Delta E_U>0$.  
This means that, in the BEC strong-coupling limit, tightly bound composite bosons gain in kinetic energy because they condense at ${\bm k}=0$ in momentum space, simultaneously when the order parameter becomes nonzero.  
The loss of the potential energy arises because the motion of composite bosons is accompanied necessarily by the breaking of on-site pairs.  
Thus, in the strong-coupling region, the superconducting state is driven by the gain in kinetic energy.  
In our calculation, the BEC features ($\Delta E_K<0$ and $\Delta E_U>0$) appear at $U/t \sim 6$.  
These behaviors of the kinetic and potential energies in the attractive Hubbard model are qualitatively consistent with the results of the previous DMFT calculations.\cite{toschi,kyung}

\section{Summary}

We have studied the superconducting ground state in the two-dimensional attractive Hubbard model by VCA.  
We have calculated the $U$ dependence of the order parameter and have shown that the order parameter is suppressed in comparison with that obtained using the mean-field theory owing to spatial fluctuations in the low-dimensional system.  
In order to discuss the character of the BCS-BEC crossover, CPT has been used to calculate the single-particle and anomalous Green's functions.  
We have shown that the single-particle spectra and densities of states clearly exhibit the behavior of the superconducting gap and that the Bogoliubov quasiparticle spectra and condensation amplitude characterize Cooper pairs in momentum space that changes continuously from the BCS state to the BEC state.  
From the calculated condensation amplitude, we have evaluated the pair coherence length $\xi$, which demonstrates the smooth crossover in real space from the weakly paired BCS state ($\xi \gg a$) to the BEC state of tightly bound pairs ($\xi \ll a$).  
We have also calculated the kinetic and potential energies in the superconducting and normal ground states and have shown that the superconducting state is driven by the gain in potential energy in the BCS state, but by the gain in kinetic energy in the BEC state.  

\begin{acknowledgments}
We thank K. Seki for useful discussions.  
T.K. acknowledges support from JSPS Research Fellowship for Young Scientists.  
This work was supported in part by the Kakenhi Grant No.~22540363 of Japan.  
\end{acknowledgments}

\end{document}